\documentclass[prl,showpacs,twocolumn,superscriptaddress]{revtex4}
\usepackage{bm,amsmath,amsfonts,paralist}
\usepackage{graphicx}
\usepackage{epstopdf}

\begin{document}

\title{Diluted one-dimensional spin glasses with power law decaying interactions}

\author{L. Leuzzi} 
\affiliation{Dipartimento di Fisica, Universit\`a di Roma ``La
  Sapienza'', P.le Aldo Moro 2, I-00185 Roma, Italy}
\affiliation{Statistical Mechanics and Complexity Center (SMC) - INFM
  - CNR, Italy}

\author{G. Parisi} 
\affiliation{Dipartimento di Fisica, Universit\`a di Roma ``La
  Sapienza'', P.le Aldo Moro 2, I-00185 Roma, Italy}
\affiliation{Statistical Mechanics and Complexity Center (SMC) - INFM
  - CNR, Italy}

\author{F. Ricci-Tersenghi} 
\affiliation{Dipartimento di Fisica, Universit\`a di Roma ``La
  Sapienza'', P.le Aldo Moro 2, I-00185 Roma, Italy}

\author{J.J. Ruiz-Lorenzo} 
\affiliation{Departamento de F\'{\i}sica, Univ. Extremadura, Badajoz,
  E-06071 and BIFI, Spain.}

\begin{abstract}
  We introduce a diluted version of the one dimensional spin-glass
  model with interactions decaying in probability as an inverse power
  of the distance.  In this model varying the power corresponds to
  change the dimension in short-range models.  The spin-glass phase is
  studied in and out of the range of validity of the mean-field
  approximation in order to discriminate between different theories.
  Since each variable interacts only with a finite number of others
  the cost for simulating the model is drastically reduced with
  respect to the fully connected version and larger sizes can be
  studied.  We find both static and dynamic evidence in favor of the
  so-called replica symmetry breaking theory.
\end{abstract}


\maketitle


Mean-field spin glass models are known to have a rather complex
low-temperature phase \cite{Parisi80}, which has not been clearly
observed so far in numerical simulations of finite-dimensional models
with short range interactions.  Theories alternative to the mean-field
one have been proposed \cite{Fisher86}, but short-range systems with
quenched disorder are very tough to study analytically
\cite{Dedominicis}.  Numerical simulations have been, thus,
extensively employed, developing more and more refined algorithms over
the years, though with no conclusive indication on the nature of the
spin-glass phase in finite dimension.

Long-range models are such that their lower critical dimension is
lower than the one of the corresponding short range model. In
particular, one can have a phase transition even in one dimensional
systems, provided the range of interaction is large enough.  One
dimensional spin glass models with power-law decaying interactions
actually allow to explore both long- and short-range regimes by
changing the power \cite{vEnter83,Kotliar85,Leuzzi99,Katzgraber03}.
These models would be perfect candidates for comparing the spin glass
phase in and out of the range of validity of the mean field
approximation.  Unfortunately, since each variable interacts with all
the others, numerical simulations are very computer demanding and it
is hard to get a clear numerical evidence supporting a specific spin
glass theory \cite{Leuzzi99,Katzgraber03}.

We, therefore, introduce a diluted version of the model, where the
mean coordination number is fixed (see also
Ref. \onlinecite{Franz06}).  In diluting, the run time grows as the
size $N$ of the system, rather than proportionally to $N^2$.  This is
a fundamental issue because finite volume effects are strong in these
models: previous studies were restricted to $N \le 512$, while we can
now thermalize systems up to $N=16384$, thus keeping these effects
under control.

We are interested in analyzing the difference among the predictions on
the spin glass phase of the droplet theory \cite{Fisher86}, the
trivial-non-trivial (TNT) scenario \cite{Krzakala00} and the replica
symmetry breaking (RSB) theory \cite{Parisi80}.  Studying the
thermodynamics, we focus on site and link overlaps, providing strong
evidence that both fluctuate in the infinite volume limit.  From the
dynamic behavior we learn that the four-point correlation function
goes to zero at large distances when extrapolated at infinite times.
In this framework we are able to identify a characteristic
length-scale $\ell(T;t)$ for such a decay.


The model investigated is a one dimensional chain of $N=L$ Ising spins
($\sigma_i=\pm 1$) whose Hamiltonian reads
\begin{equation}
{\cal H}=-\sum_{i<j} J_{ij} \sigma_i\sigma_j\;.
\label{eq:ham}
\end{equation}
The quenched random couplings $J_{ij}$ are independent and identically
distributed random variables taking a non zero value with a
probability decaying with the distance between spins $\sigma_i$ and
$\sigma_j$, $r_{ij}=|i-j| \mod (L/2)$, as
\begin{equation}
\mathbf{P}[J_{ij}\neq 0] \propto r_{ij}^{-\rho}\;.
\label{eq:Jij}
\end{equation}
Non-zero couplings take value $\pm 1$ with equal probability.  We
choose an average coordination number $z=6$ and periodic boundary
conditions.

\begin{table}[b!]
\begin{tabular}{|l|l|}
\hline
$\rho < 1$ &	 Bethe lattice like\\
$1<\rho\leq 4/3$ & $2^{\rm nd}$ order transition, mean-field (MF)\\
$4/3<\rho<2$	 & $2^{\rm nd}$ order transition, infrared divergence (IRD)\\
$\rho=2$ & Kosterlitz-Thouless or $T=0$ phase transition\\
$\rho>2$ & no phase transition\\
\hline 
\end{tabular}
\protect\caption{From infinite range to short range behavior of the
  spin-glass model defined in Eqs.(\ref{eq:ham},\ref{eq:Jij}).}
\label{tab:rho}
\end{table}

The universality class depends on the value of the exponent $\rho$,
and it turns out to be equal to the one of the fully connected version
of the model, where bonds are Gaussian distributed with zero mean and
a variance depending on the distance as ${\overline{J_{ij}^2}} \propto
r_{ij}^{-\rho}$~\cite{vEnter83,Kotliar85,Leuzzi99,Katzgraber03}.  The
overline denotes the average over quenched disorder.

As $\rho$ varies this model is known to display different statistical
mechanics behaviors. For the diluted case they are reported in
Tab.~\ref{tab:rho}.  In the limit $\rho\to 0$ the model is a
spin-glass on a Bethe lattice \cite{Viana,Mezard00}, at variance with
the fully connected version where this limit is ill-defined for any
$\rho<1$.  If the decay is gentle enough ($\rho \le 4/3$), the
mean-field (MF) approximation is exact.  As it becomes steeper ($\rho
> 4/3$), the MF approximation breaks down because of infrared
divergences (IRD).  For $\rho=2$ one does not expect a finite
temperature phase transition, though power law correlations might
still be there \footnote{In long-range systems with no quenched
  disorder the equivalent of this model case displays a
  Kosterlitz-Thouless  transition \cite{Kosterlitz76}.}.  This
special case deserves further investigation.

The $\rho=4/3$ case corresponds to the upper critical dimension of
short-range spin-glasses in absence of an external magnetic field
($D=6$), whereas $\rho=2$ plays the role of the lower critical
dimension. An approximate relationship between $\rho$ and the
dimension $D$ of short-range models can be identified as follows.  In
long-range models, the free theory in the replica space reads
\cite{Leuzzi99}
\begin{equation}
{\cal H}=\frac{N}{4}
\int\frac{dk}{2\pi}
\left(k^{\rho-1}+m_0^2\right)
\sum_{a\neq b}\left|\tilde Q_{ab}({\bm k})\right|^2\;,
\label{f:free_th_fc}
\end{equation}
where $a$ and $b$ are replica indices and $\tilde Q_{ab}(\bm k)$ is
the Fourier transform of the distance-dependent overlap matrix element
$Q_{ab}(r_{ij})$.  Comparing the critical scaling ($m_0 \propto
|T-T_c|=0$) of Eq.(\ref{f:free_th_fc}) with that of the free theory
for short-range spin glass models in $D$ dimensions ($\mathcal{H} \sim
\int d^Dk\;k^2 {\rm Tr}\,Q^2$) the following equation turns out to
hold close to the upper critical dimension
\begin{equation}
\rho=1+2/D \, .
\end{equation}

\begin{figure}
\includegraphics[width=.43\textwidth]{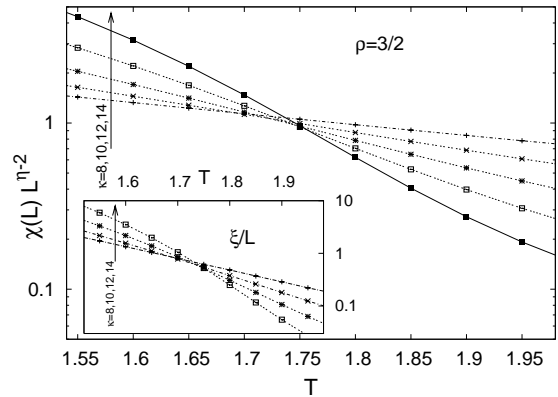}
\caption{$\rho=3/2$, IRD regime. Plot of $L^{\eta-2}\chi_{\rm sg}$
  vs.\ $T$. Inset: $\xi_L/L$ vs.\ $T$. Sizes are $L=2^\kappa$, with
  $\kappa=8,10,12,14$.}
\label{fig:FFS_Tc_rho15_chi}
\end{figure}

We study the equilibrium properties of the diluted long-range model
both in ($\rho=5/4$) and out ($\rho=3/2,5/3$) of the MF regime for
sizes up to $L=16384=2^{14}$.  We simulate two replicas
$\sigma_i^{(1,2)}$ using the parallel tempering algorithm \cite{PT}
with 20 temperatures and we measure site and link overlaps,
respectively defined as
\begin{equation}
q = \frac{1}{N} \sum_{i=1}^N \sigma_i^{(1)}\sigma_i^{(2)}\;,
\quad
q_l = \frac{1}{z N}\sum_{i,j}^{1,N} J_{ij}^2 \sigma_i^{(1)}\sigma_j^{(1)}
\sigma_i^{(2)}\sigma_j^{(2)}\;.
\end{equation} as well as the correlation length~\cite{Cooper89}:
\begin{equation}
\xi_L=\frac{1}{2\sin(\pi/L)}\left[\frac{\chi_{\rm sg}}{\tilde
\chi(2\pi/L)}-1\right]^{\frac{1}{\rho-1}}\;,
\end{equation}
where $\chi_{\rm sg}=L{\overline{\left<q^2\right>}}$ is the spin glass
susceptibility ($\langle(\cdots)\rangle$ denotes the thermal average
and $\overline{(\cdots)}$ denotes the average over the disorder) and
$\tilde\chi(k)$ is the Fourier transform of the four-point correlation
function ($\tilde\chi(0)=\chi_{\rm sg}$). Averages over the disorder
are taken on $O(3\times 10^5)$ samples in the smallest lattices and
over $O(2\times 10^4)$ samples in the larger ones.  In order to
compute critical temperatures, critical exponents and the Finite Size
Scaling (FSS) corrections we have used the quotient
method~\cite{Ballesteros98}.  We have computed the exponent $\nu$
using the scaling of the temperature derivative of $\xi_L/L$ and
$\eta$ from the scaling of $\chi_{\rm sg}$.  As a typical case, we
show in Fig.~\ref{fig:FFS_Tc_rho15_chi} the temperature and size
dependence of $\chi_{\rm sg}$ and $\xi_L$.  In the quotient method the
estimates of the critical exponent still depend on the lattice size:
the extrapolation to infinite volume provides both their asymptotic
values and the $\omega$ exponent of the leading FSS correction,
$O(L^{-\omega})$.

\begin{table}[t!]
\begin{tabular}{l|c|c|c|c|c|c|}
\cline{2-7}
\hspace*{.8cm} & $\quad\rho\quad$& $\,T_c\,$      & $\,1/\nu\,$ & $\eta$ & $\,\eta$ (th.)~ & $\omega$ \\
\hline
MF & $5/4$ & $2.191(5)$ & $0.28(2)$ & $1.751(8)$   & $1.75$ & $0.40(2)$\\
IRD & $3/2$ & $1.758(4)$ & $0.25(3)$ & $1.502(8)$   & $1.5$  & $0.60(6)$\\
IRD & $5/3$ &   $1.36(1)$ &  $0.19(3)$ & $1.32(1)$   & $1.3{\bar 3}$& $0.8(1)$\\
\hline
\end{tabular}
\protect \caption{Estimates of critical temperature and exponents.}
\label{tab:critical}
\end{table}

The results are summarized in table \ref{tab:critical}. The $\eta$
exponent coincides with the theoretical prediction $\eta=3-\rho$
($\eta$ is not renormalized in the IRD regime
\cite{Kotliar85,Leuzzi99}).  Due to strong finite size effects this
check failed in previous works \cite{Katzgraber03}.  The $\nu$
exponent is consistent with the theoretical prediction,
$\nu=1/(\rho-1)$, in the MF case. In the IRD regime, thermodynamic
fluctuations dominate and a renormalization is necessary: at present
only one loop calculations are available \cite{Kotliar85,Leuzzi99},
but the estimate of $\nu$ is too rough to compare with numerical data.

In the spin glass phase ($T<T_c$), site and link overlap
distributions, $P(q)$ and $P_l(q_l)$, can be used as hallmarks to
discriminate among different theories for finite dimensional spin
glasses.  Indeed, three cases are contemplated in the literature.
\begin{compactenum}
\item Droplet theory: one state; both distributions delta-shaped.
\item TNT scenario: many states ($q$ fluctuates), but droplet-like
  excitations ($q_l$ fluctuations vanish for large sizes); non-trivial
  $P(q)$ and trivial $P_l(q_l)$.
\item RSB theory: many states with space-filling excitations; both
  distributions broad.
\end{compactenum}

\begin{figure}
\includegraphics[width=.43\textwidth]{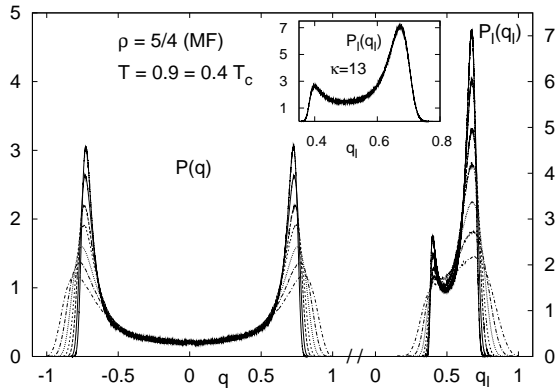}
\caption{$P(q)$ and $P_l(q_l)$ at $\rho=5/4$ (MF). $T=0.9\simeq 0.4
  T_c$, with $L=2^\kappa$ and $\kappa=7,\ldots,13$. Inset: $P_l(q)$
  for $L=2^{13}$.}
\label{fig:Pq_rho125}
\end{figure}

\begin{figure}
\includegraphics[width=.43\textwidth]{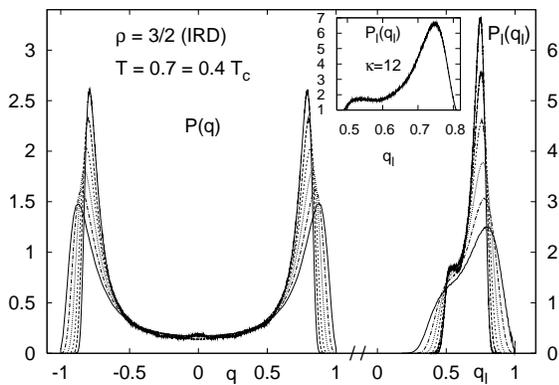}
\caption{$P(q)$ and $P_l(q_l)$ at $\rho=3/2$ (IRD). $T=0.7\simeq 0.4
  T_c$, with $L=2^\kappa$ and $\kappa=7,\ldots,12$. Inset: $P_l(q)$
  for $L=2^{12}$.}
\label{fig:Pq_rho15}
\end{figure}

Distributions $P(q)$ and $P_l(q_l)$ for $T\simeq 0.4 ~T_c$ are plotted
in Figs.~\ref{fig:Pq_rho125} and \ref{fig:Pq_rho15} in a model case
where MF is exact ($\rho=5/4$) and in a IRD case ($\rho=3/2$),
respectively.  In both cases, we see two peaks in the $P_l(q_l)$ for
large sizes (insets).  Such a result would have been impossible to
observe out of MF with sizes smaller than $L=2^{12}$ \footnote{To
  reach low temperatures ($T\simeq 0.4 T_c$) we have performed
  additional PT runs using $40$ temperatures, simulating a smaller
  number of samples ($O(10^3)$) for sizes up to $L=2^{13}$ (for $\rho=5/4$) and
$L=2^{12}$ (for $\rho=3/2$).}.

\begin{figure}
\includegraphics[width=.43\textwidth]{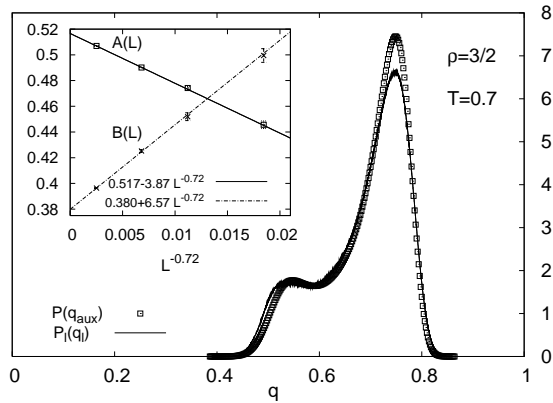}
\caption{Distributions of $q_l$ (line) and $q_{\rm aux}$ (empty
  squares) for $\rho=3/2$, $L=2^{12}$ and $T=0.7$. Inset: A and B
  vs. $L^{-0.72}$ obtained by measuring the Kullback-Leibler
  divergence between the two distributions for $L=2^\kappa$,
  $\kappa=6,8,10,12$.}
\label{fig:2distr}
\end{figure}

Both distributions seem to be broad, but their thermodynamic limits
must be carefully controlled.  While it is easy to convince oneself
that $P(q)$ is not a bimodal distribution as $L\to\infty$ (e.g.,
$P(0)$ is practically size independent), the limit of $P_l(q_l)$ is
more difficult to extract from finite size data, since its variance
converges to a small value.  We provide, thus, an alternative method
of analysis testing the hypothesis that both $q$ and $q_l$ are
equivalent measures of the distance among states \cite{Athanasiu87}.
The simpler relation between the two overlaps is
\begin{equation}
q_l \stackrel{{\rm d}}{=} q_\text{aux}\equiv A + B q^2 + C \sqrt{1-q^2} z\;,
\label{eq:trasf}
\end{equation}
where $z$ is a Gaussian random variable with zero mean and unitary
variance mimicking finite size effects, and $A$, $B$ and $C$ are
fitting parameters.  Such a relation is satisfied in the
Sherrington-Kirkpatrick model, with $A=C=0$, and it is a good
approximation for the short-range spin glass in $D=3$
\cite{Hed07,Contucci07}.  For each value of $L$, at $\rho=3/2$ and
$T=0.7$, we compute the best fitting parameters by minimizing the
symmetrized Kullback-Leibler (KL) divergence \cite{Kullback51} between
the distribution of $q_l$ and that of $q_\text{aux}$.  In
Fig.~\ref{fig:2distr}, for $L=2^{12}$, we compare the optimal
distributions, which should coincide if Eq.~(\ref{eq:trasf}) held.
Eq.~(\ref{eq:trasf}) provides a strong evidence for a non-trivial link
overlap distribution as long as the $B$ parameter converges to a non
zero value for large size, as one can verify in the inset of
Fig.~\ref{fig:2distr} where we show such an extrapolation for $A$ and
$B$ plotted vs.\ an inverse power of $L$ ($C$ and the optimal KL
divergence go to zero, as expected).

\begin{figure}
\includegraphics[width=.4\textwidth]{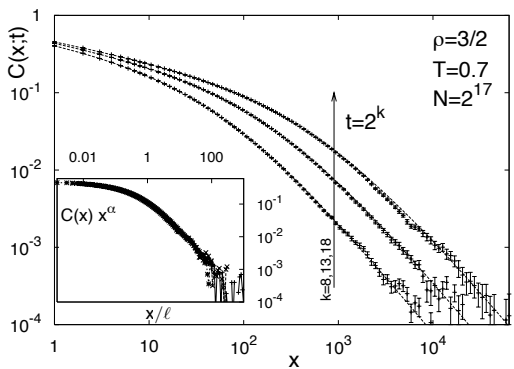}
\vskip -1mm
\caption{$C_q(x,t)$ for $\rho=3/2$, $L=2^{17}$ at $T=0.7\sim 0.4T_c$
  and different times $t=2^{8},2^{13},2^{18}$. The curves are fits to
  Eq.(\ref{f:Cx_fit_t}).  Inset: data collapse by plotting $C(x;t)~
  x^\alpha$ vs.\ $x/\ell(t)$, for all times between $2^8$ and
  $2^{18}$.}
\label{fig:qq_dyn_rho15}
\end{figure}
\begin{figure}
\includegraphics[width=.4\textwidth]{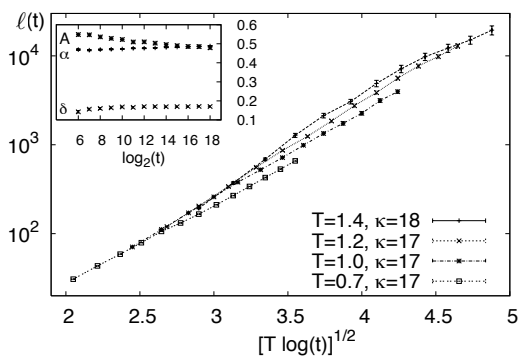}
\vskip -2mm
\caption{Crossover length $\ell(t)$ vs.\ $\sqrt{T\log t}$ for
  $\rho=3/2$, $L=2^\kappa$ and $T$ between $0.7 \simeq 0.4 T_c$ and
  $1.4\simeq 0.79 T_c$. Inset: parameters $A$, $\alpha$ and $\delta$
  vs.\ time at $T=0.7$.}
\label{fig:fitpar}
\end{figure}

As a complementary approach we look at the off-equilibrium four-point
correlation function
\begin{equation}
C_q(x,t)=\frac{1}{L}\sum_{i=1}^L{\overline{\langle\sigma^{(1)}_i(t)\sigma^{(2)}_i(t)
\sigma^{(1)}_{i+x}(t)\sigma^{(2)}_{i+x}(t)\rangle}}\;.
\end{equation}
For very large distances the fastest decay expected goes like
$x^{-\rho}$, because of long-range interactions.  For intermediate
distances, up to an effective crossover length $\ell(t)$, we observe a
slower decay $x^{-\alpha}$, with $0 < \alpha < \rho$, which is
incompatible with the onset of a plateau at $q_{\rm EA}^2$ in the
large times limit.  This suggests to use the function
\begin{equation}
A\,x^{-\alpha} \left[
1+\left(\frac{x}{\ell}\right)^{\delta(\rho-\alpha)}
\right]^{-1/\delta}
\label{f:Cx_fit_t}
\end{equation}
to interpolate $C_q(x,t)$ data at a fixed time $t$.  The fits are very
good and their quality can be appreciated in
Fig.~\ref{fig:qq_dyn_rho15} for an IRD ($\rho=3/2$) system of size
$L=2^{17}$.  The crossover length $\ell$ plays a role similar to the
correlation (or coherence) length in short range spin glasses
\cite{Kisker96,Marinari00,Berthier02}.  We allow the fitting
parameters $A,\alpha$ and $\delta$ to depend on time.  Nonetheless we
observe (see inset of Fig. \ref{fig:fitpar}) that they become
stationary for large times: this is a strong evidence that
Eq.~(\ref{f:Cx_fit_t}) is a significant and robust behavior.

The growth of $\ell(t)$ with time at different temperatures below
$T_c$ is plotted in Fig.~\ref{fig:fitpar}. The length $\ell(t)$
reaches very large values ($> 10^4$) with respect to previously
studied spin glass models \cite{Kisker96,Marinari00,Berthier02}.  In
this region $\ell(t)$ is very well fitted by the phenomenological law
$a(T) \exp(b(T) \sqrt{T \log t})$, with $a$ and $b$ not very dependent
on the temperature; this seems reasonable since in activated processes
the typical scaling variable is $T \log(t)$.

We also tried to fit the previous $\ell(t)$ data with a generalized
droplet scaling \cite{Berthier02} $\tau(\ell) = A(T) \ell^{z_c}
\exp(\Upsilon(T)\ell^\psi)$, where the power-law factor dominates near
the transition ($\lim_{T\to T_c}\Upsilon(T)=0$) and the exponential
term governs the low temperature regime.  The critical exponents $z_c$
and $\psi$ are predicted not to depend on $T$.  The data shown in
Fig.~\ref{fig:fitpar} are not compatible with this scaling law for any
temperature-dependent $A(T)$ and $\Upsilon(T)$.

In conclusion, we have introduced a model which is easy to simulate
and allow to probe the spin glass phase beyond mean-field.  In this
regime, from the analysis of thermodynamics, we observe that both the
site and the link overlap parameter fluctuate for large sizes.  In the
large times limit the out-of-equilibrium four-point function
$C_q(x,t)$ tends to a well defined function that displays a power-law
decay to zero and is incompatible with the onset of a plateau at any
large $x$.  These observations are consistent with the clustering
properties of the RSB theory.  The bond diluteness of the model under
investigation strongly reduces simulation times and allows to
thermalize systems of sizes large enough to clearly discern the double
peak structure of $P_l(q_l)$.  Both droplet and TNT proposal appear,
in conclusion, not consistent with a FSS analysis over large sizes and
with the behavior of the four-point correlation function and the
related coherence length.  This model is well suited to study the spin
glass transition in a magnetic field and work is in progress in this
direction.

We thank Silvio Franz for useful discussions and suggestions.  This
work has been partially supported by MEC, contracts FIS2006-08533-C03
and FIS2007-60977. Part of simulations were performed in the BIFI
cluster.

\end{document}